\documentclass[fleqn,twoside]{article}
\usepackage{espcrc2}

\usepackage{graphicx}
\usepackage[figuresright]{rotating}

\newcommand{\AmS}{{\protect\the\textfont2
  A\kern-.1667em\lower.5ex\hbox{M}\kern-.125emS}}

\title{Confining properties of a gas of $Z(2)$ vortices}

\author{Srinath Cheluvaraja \address{Department of Physics and
Astronomy, Louisiana State University, 
        \\ 
        Baton Rouge, USA }%
        \thanks{ This research was supported in part by United States
Department of Energy grant DE-FG 05-91 ER 40617.
                }}
       
\begin{document}
\def \beq{\begin{equation}}
\def \eeq{\end{equation}}
\def \beqa{\begin{eqnarray*}}
\def \eeqa{\end{eqnarray*}}
\mathindent=0pt
\begin{abstract}
Vortex solutions are studied in an SO(3) gauge theory spontaneously
broken to SO(2). These vortices have a Z(2) magnetic charge.
A dilute gas of Z(2) vortices is studied taking into
account vortex-vortex interactions. By going to a dual representation
we show that odd charges are confined with a string tension which
decreases exponentially with the inverse coupling.
\vspace{1pc}
\end{abstract}

\maketitle

\section{Introduction}

Topologically non-trivial solutions are known to affect the long
distance properties of gauge theories. By expanding around a gas of
such solutions
the
partition function of gauge theories can be rewritten using a dual
representation.
The case of
compact electrodynamics in a spontaneously broken gauge
theory was analysed by Polyakov \cite{poly77} where it was shown that 
a dilute gas of monopoles
produces confinement of charges in the Higgs phase.
 Recently there has been a lot of discussion
on $Z(2)$ vortices as a possible disordering mechanism for Wilson
loops. Since $Z(2)$ vortices can emerge as classical solutions in gauge
theories it is of interest to see if a gas of such $Z(2)$ vortices can
be analysed in the same fashion.
Classical solutions with $Z(2)$ magnetic charge are  also known in
cosmology and are called cosmic strings \cite{vacha90}. However, cosmic strings have the
property that a particle going round it becomes its anti-particle. Such
cosmic strings are associated with a loss of local charge conservation.
The solutions we will consider are not cosmic strings but more like
Abrikosov-
Nielsen- Olesen vortices in
a superconductor \cite{abri56,niels73}.

\section{Z(2) Vortices}

The field theory under consideration is an $SO(3)$ gauge invariant theory
minimally coupled to a matrix valued Higgs field $M$ in the $(3,3)$
representation of $SO(3)$. By studying an $SO(3)$ theory 
we would like to highlight the fact that the vortices are not
in any way dependent on the center of the gauge group but are a consequence
of the topology of the group.
The Lagrangian (in $2+1$ space-time dimensions) is given by
\beqa
L= \int d^{2}x\ d\tau   \frac{1}{2} tr(D_{\mu}M)^{t}(D_{\mu}M) - \\
\frac{1}{4}F_{\mu \nu}^{\alpha}
F_{\mu \nu}^{\alpha} -tr V(M^{t}M) 
\quad ,
\eeqa
with the usual notations.
Under local $SO(3)$ transformations ($V(x)$) 
the fields transform as
\begin{eqnarray*}
M\rightarrow VMV^{-1} \\
A_{\mu} \rightarrow V A_{\mu} V^{-1} -\frac{1}{g}(\partial_{\mu}V)V^{-1}
\quad .
\end{eqnarray*}

The Higgs potential is chosen to be
\beq
V(M)= \lambda (M^{t}M-I)^2
\quad ;
\eeq
this choice breaks the gauge symmetry to $SO(2)$.
Vortex solutions have the following form for the gauge fields and
the scalar field,
$A_{0}^{\alpha}=0,
A_{\mu}^{1}=A_{\mu}^{2}=0,
A_{\theta}^{3}= A(r),
A_{r}^{3}=0,
M(r,\theta)=M_{c}(r)M(\theta)$.
The equations of motion become
\begin{eqnarray*}
D_{i}F_{i j \alpha} =g\ tr((D_{j}M)^{t}[T^{\alpha},M]) \\
\partial_{i}(D_{i}M)=-g[A_{i},D_{i}M] +\frac{\partial V}{\partial M}
\quad .
\end{eqnarray*}
If we look for solutions in the type II limit ($m_{S} >> m_{V},
m_{V}=\alpha g$)
the equation for the gauge field becomes
\beq
\frac{d^{2} A(r)}{d r^2} + \frac{d}{d r}(\frac{A(r}{r})
=\alpha^2(A(r) g^2 -\frac{g}{2 r})
\quad ,
\eeq
where $\alpha^2=tr([T^3,M(0)]^{t}[T^3,M(0)])$. 
\section{Dual representation}

The partition function of a dilute gas of such vortices can be written
as
\beq
Z=\sum_{N}\frac{\mu^N}{N!}\int\prod_{j=1}^{N}dR_{j}
\exp[-\frac{\pi m_{V}^2}{4 g^2}\sum_{a\ne b}K_{0}(m_{V}|R_{a}-R_{b}|)]
\quad .
\eeq
The chemical potential term contains the self-energy of a single vortex.
Once we have this expression we can use
the method of Polyakov as used in \cite{schap78} in the analysis of
a gas of Abrikosov vortices and go over to the dual field 
$\chi(x)$. 
This again results in a sine-gordon
theory

\beq
Z=\int D\chi \exp \ -\frac{A}{2}\int [(\nabla \chi)^2 +m_{V}^2 \chi^2
-M^2 Cos(\chi)]d^{2}x\
\quad ,
\eeq
where
$A=\frac{g^2}{\pi^2m_{V}^2}$,
$M^{2}=\frac{2\pi^2 m_{V}^2}{g^2} \exp(-\epsilon_{1})$, $\epsilon_{1}$
is the energy of a single vortex.
The correlation functions in a gas of $Z_{2}$ vortices can be calculated
by defining a generating function
\beq
<\exp i \ \int \eta(x) \rho(x) d^2 x> = \frac{Z[\eta]}{Z[0]}
\quad,
\eeq
$\rho(x)=\sum_{a}\delta(x-x_{a})$.

and $Z[\eta]$ is given by
\beqa
Z=\int D\chi \exp \ \frac{-A}{2}\int [(\nabla (\chi - \eta))^2 +m_{V}^2
(\chi - \eta)^2
\\
-M^2 \cos \chi ]d^{2}x\
\quad .
\eeqa

The Wilson loop
\beq
\exp q \int_{C}  A_{\mu}dx^{\mu}
\quad ,
\eeq
becomes
$\exp \pi q T^{3}$ if it encircles a $Z(2)$ vortex and has trace $-1$ for
odd multiples of $g$. In the absence of a vortex the trace is $+3$.

There is an old argument which shows that in a gas of randomly distributed
magnetic fluxes the Wilson loop will have an area law and this argument
can be used to show that a gas of $Z(2)$ vortices confines charges. However,
we can  get a more quantitative estimate for the Wilson loop taking into account
vortex-vortex interactions
by going to the dual representation.
Using the properties of $SU(2)$ and $SO(3)$ representations we can write
\beqa
(tr\ \exp q \int_{C} A_{\mu}^{3}T^{3}dx_{\mu})-1 
\\
= tr\ (\exp i (q\ \int_{C} A_{\mu}^{3}
\frac{\tau^3}{2}dx_{\mu})^2
\quad ,
\label{su2so3}
\eeqa
and define the normalized Wilson loop 
\beq
W_{n}(C)=\frac{1}{2} (tr\ \exp q \int_{C} A_{\mu}^{3} T^{3}dx_{\mu}-1)
\eeq
which takes values from $-1$ to $+1$.
The expectation value of the normalized Wilson loop can be written as
(using \ref{su2so3})
\beq
W_{n}(C)=tr\ <\exp i \int 2 q \rho(x) \frac{\tau^3}{2}\eta(x)d^{2}x>
\quad ,
\eeq
where $\eta(x)$ is given by
\beq
\eta(x)=\frac{m_{V}^{2}}{2 g}\int_{S_{C}} K_{0}(m_{V}|x-x'|)dx'
\label{eta}
\quad .
\eeq
$S_{C}$ is the minimal surface spanning the loop $C$.
Effectively we have to calculate the average value of a diagonal
matrix whose elements are
\beq
W_{n}(C)=<\exp i \int q \rho(x) \eta(x)d^{2}x>
\quad .
\eeq
This can be done by using the method of \cite{poly77} and
\cite{schap78} and we repeat the main steps here.

Using the dual representation
\beq
W_{n}(C)= \frac{Z_{1}}{Z_{2}}
\quad ,
\eeq
where the numerator and denominator are given by
\beq
Z_{1}=\int D\chi \exp -\frac{A}{2}\int [(\nabla (\chi - \eta))^2 +m_{V}^2
(\chi - \eta)^2
-M^{2} \cos \chi ]d^{2}x\
\quad ,
\eeq
\beq
Z_{2}=\int D\chi \exp -\frac{A}{2}\int [(\nabla \chi )^2 +m_{V}^2 \chi^2
-M^{2} \cos \chi ]d^{2}x\
\quad .
\eeq
Since $\eta$ vanishes outside the surface the numerator and denominator
are the same outside the surface $S_{C}$ and have to be calculated only
on the surface $S_{C}$.
The denominator is
\beqa
\int D \chi \exp \ -\frac{A}{2}\int_{S_{C}} [-\chi (\nabla^2) \chi+ m_{V}^2
\chi^2 \\
-M^2 \cos(\chi)]d^{2}x\ \\
\quad .
\eeqa
The numerator can be split into two pieces (after an integration by
parts)
\beqa
\int D \chi \exp \ -\frac{A}{2}\int_{S_{C}} [-\chi (\nabla^2) \chi+ m_{V}^2
\chi^2
-M^2 \cos(\chi)]d^{2}x\ \\
\exp \ -\frac{A}{2}\int_{S_{C}} [(\nabla^2 \chi - m_{V}^2 \chi
+\frac{\pi m_{V}^2 q}{g}) \eta -\frac{\pi m_{V}^2 q}{g}\chi]
\eeqa
In the mean field approximation ( good in the weak coupling limit)
the integrals over the numerator and the
denominator 
yield the following equations
\beqa
\nabla^{2}(\chi_{n}-\eta)=\frac{M^2}{2} \sin \chi_{n} + m_{V}^2(\chi_{n}-\eta) \\
\nabla^{2}(\chi_{d})=\frac{M^2}{2} \sin \chi_{d} + m_{V}^2(\chi_{d})
\quad .
\eeqa

Using
\beq
(\nabla^2 - m_{V}^2) K_{0}(m_{V}r)=-2\pi\delta(r)
\eeq
and Eq.~ \ref{eta} the first equation becomes 
\beq
\nabla^2 \chi_{n} = M^2 \sin(\chi_{n}) +m_{V}^2(\chi_{n}
-\frac{\pi q}{ g} \Theta(r,S_{C}))
\quad ,
\eeq
$\Theta(r,S_{C})$ is the surface delta function
\beqa
\Theta(r,S_{C})=1 \quad if \quad  r \ \in S_{C} \\
\Theta(r,S_{C})=0 \quad otherwise
\quad .
\eeqa
The mean field equations are solved by the constant solutions
$\chi=-(2 n+1) \pi$ for $q=(2 n+1)g$ and
$\chi=-(2 n) \pi$ for $q=(2 n)g$. Substituting these constant solutions
in the numerator and the denominator we get
\beqa
W_{n}(C)=1 \quad for \ q=(2 n)g \\
W_{n}(C)= \exp ( -\sigma S_{C}) \quad  for q=(2 n +1)g \\
\quad .
\eeqa
The string tension $\sigma$ is exponentially small in the coupling
$\sigma = A M^2 \approx \exp(-\frac{1}{g^2})$.

So far our analysis has been completely classical. We have rewritten the
partition function in the Higgs phase as a gas of $Z_{2}$ vortices.
Quantum effects are included by expanding about the classical
solution and integrating out the fluctuations. This is a standard
procedure \cite{poly77} and the integration of the zero modes
renormalizes the chemical potential
to
$\mu = \frac{1}{g^{3}} f(\lambda) \exp (-\frac{\epsilon}{g^2})$ 
where $f(\lambda)$ is a determinant 
function which can in principle be computed.
For weak coupling the quantum fluctuations donot change the classical
results in any essential way.

\section{Conclusions}
In the above analysis we have classical Z(2) vortex solutions
occurring in certain broken gauge theories. The properties of
a gas of such vortices taking into account their mutual interactions
can be studied by going to a dual representation using the methods in
\cite{poly77,schap78}. This analysis yields a quantitative expresession
for the Wilson loop and odd charges are confined while even charges are
not confined. The semi-classical analysis given here presents an
instructive example where $Z(2)$ vortices confine charges.

\end{document}